\documentclass[final,1p,times]{elsarticle} % do not change
\usepackage{graphicx} % do not change
\usepackage{amssymb} % do not change
\usepackage{amsthm} % do not change
\usepackage{lineno} % do not change

\journal{Nuclear Physics A} % do not change
\begin{document} % do not change

\begin{frontmatter} % do not change

%% QM09Author: please enter your  
%% Title, author and address info here; please do not use footnotes

% Your Title - please modify
\title{Ridge, Bulk, and Medium Response:  How to Kill Models and Learn Something in the Process}

% Principle author, and co-authors - please modify
\author{J.L. Nagle}% etc

% Address - please modify
%\address[a]{University of Colorado,
\address{University of Colorado, Boulder, CO 80309-0390, USA}

\begin{abstract} % do not change
%% Text of abstract goes here - please modify
In these proceedings, we highlight experimental data (published and preliminary)
related to jet quenching and the response of the medium to this deposited energy.  
Signatures in two- and three- particle 
hadron correlations indicate interesting structures near the trigger particle in azimuth and over a broad
range in pseudo-rapidity, often termed 'the ridge', and conical-like structures separated in azimuth 
opposite to the trigger particle.  
We review numerous theoretical interpretations of the ridge in particular with a critical eye for the
key properties that allow one to discriminate between, or rule out, certain physical pictures and models
(and hopefully learn something in the process).
\end{abstract} % do not change

%\begin{keyword} % do not change
% please check/modify - you may not have the same physics topic
%\PACS 25.75.Dw % Particle and resonance production in heavy ion collisions
%\end{keyword} % do not change

\end{frontmatter} % do not change

%% QM09: we keep linenumbers at least for initial version
%\linenumbers % do not change

%% start of main text - please modify. Below is a sub-set (single section) 
%% of an earlier proceedings that show how one can handle references 
%% and figures etc.
%%\section{}\label{}
\section{What the Parton Deposits in the Medium}

A necessary prerequisite for understanding any medium response to the traversal of a high-energy
parton is to understand in detail how the parton deposits energy (i.e. how much energy on average and
with what distribution in energy and in space-time).  The measurement of single inclusive high $p_T$ hadron
suppression ($R_{AA}$) has proven to be very useful for making an overall characterization of the energy loss. 
However, there are a variety of parton energy-loss formalisms that currently give approximately equal 
goodness-of-fit to the experimental data, but with very different implications
for the quenching power of the medium~\cite{ppg079,ppg080}. To be clear, the $R_{AA}$ comparison allows one to
tightly constrain this quenching power within a given model, but not to discriminate between models.

\begin{figure}
\centering
\includegraphics[scale=0.39]{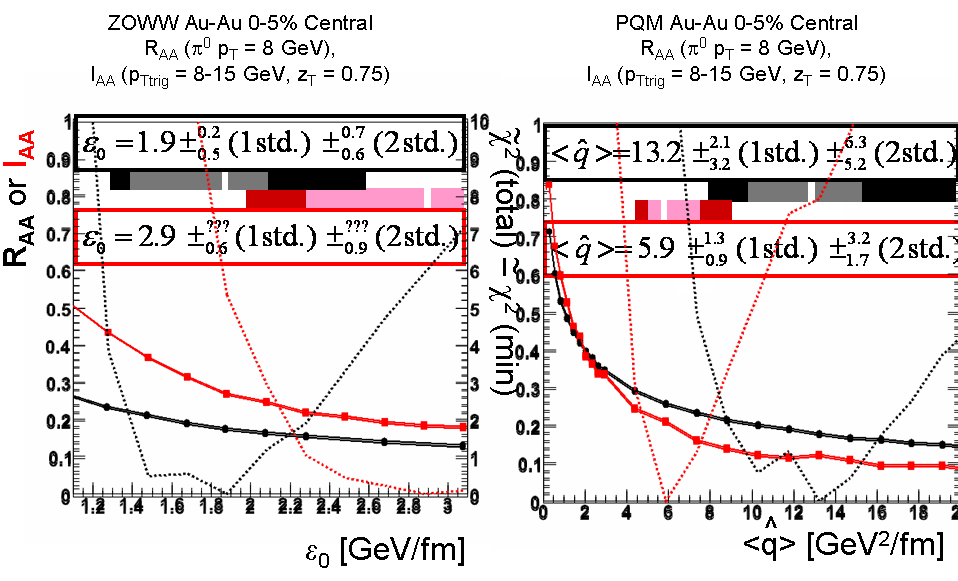}
\caption{(color online) Shown are the results of calculations from ZOWW~\cite{zoww} (left panel) and PQM~\cite{pqm} (right panel) 
for $R_{AA}$ (black solid line) and $I_{AA}$ (red solid line) 
(left y-axis) as a function of $\epsilon_{0}$ or $\left< \hat q \right>$.  
Also shown (right y-axis) are $\tilde \chi^2 - \tilde \chi^{2}_{minimum}$ for $R_{AA}$ (black dashed line) and for $I_{AA}$ (red dashed line)
when comparing theory to experimental data~\cite{ppg079,ppg080,star_dihadron}.
}
\label{fig_constraint}       % Give a unique label
\end{figure}

It has been proposed that di-jet (as well as direct photon-jet and fully reconstructed jet) observations will provide this discriminating power.
First measurements via di-hadron correlations at high $p_T$ have been made by the STAR experiment, though they are
somewhat statistics limited in the $d-Au$ baseline measurement~\cite{star_dihadron}.  In the context of two 
parton energy-loss models (ZOWW~\cite{zoww} and PQM~\cite{pqm}), we show in Figure~\ref{fig_constraint} the calculation results
for $R_{AA}$ and $I_{AA}$ as a function of the quenching power (either $\epsilon_{0}$ or $\left< \hat q \right>$).  
Also shown are the resulting modified $\tilde \chi^{2} - \tilde \chi^{2}_{minimum}$ (including experimental statistical and systematic uncertainties, 
but not theoretical uncertainties) when confronted with the $R_{AA}$ and $I_{AA}$ data.  
It is striking that the best fit is a larger (smaller) quenching implied from $I_{AA}$ as opposed to $R_{AA}$ for ZOWW (PQM) model.
There are also recent results with ASW energy-loss embedded in a hydrodynamical medium that result in a
somewhat more consistent description of both $R_{AA}$ and $I_{AA}$~\cite{spain_eloss}.  
%The probability of two measurements are consistent with the same quenching in these models, but the differences
%are the results of experimental statistical fluctuations is equivalent to both values being 
%offset from the true value by 1.5 $\sigma$ (or more) and is less than 1.7\%.

This provokes one to question what has been learned since this was a more discriminating observation.  
We can speculate that in a model with the same average energy loss, but smaller fluctuations (i.e. a larger number
of lower energy radiated gluons), one might have more surface bias and thus
a lower $I_{AA}$ for the same $R_{AA}$.  Currently other model differences such as medium geometry make
it challenging to discriminate between the underlying energy-loss physics mechanisms. 
The TECHQM effort~\cite{techqm} is working to resolve these issues, but it is undermined if the community
simply claims that ``all models roughly agree with the data.''
The resolution to discriminating between the parton energy-loss 
formalisms will not come exclusively from theory; and there are 
additional experimental handles, including as an example the recent measure of $R_{AA}$ versus reaction plane~\cite{rwei}.  

\section{The Ridge}

The ridge is defined as a correlation in momentum space between particles that are extended over many
units in pseudo-rapidity ($\Delta \eta$) and relatively narrow in azimuthal angle ($\Delta \phi$).  In fact, there are
multiple observations of 'ridge-like' phenomena (which may or may not originate from a common physics mechanism).  
There are preliminary results of a 'hard ridge' where a high $p_T$ trigger particle (3-12 GeV/$c$) is correlated with a 
moderate $p_T$ associated particle (2-4 GeV/$c$)~\cite{putschke}.  There are also preliminary results of a 'soft ridge' where
the two correlated particles are simply required to have $p_{T} > 0.150$ GeV/$c$~\cite{daugherity}.  In between the two (which
was referred to as the 'just right ridge') are published results from the PHOBOS experiment with trigger particles with $p_{T} > 2.5$ 
GeV/$c$ and associated particles with $p_{T} > 0.035$ GeV/$c$~\cite{phobosridge}.

Preliminary studies from PHENIX and STAR of the transverse momentum spectra, hadron chemistry (i.e. baryon/meson ratios), 
and centrality scaling have been shown~\cite{putschke,jchenphenix}.  These reveal that the $p_T$ spectra in the ridge is similar to that
of the bulk medium (with some indications of being slightly harder).  This is in striking contrast to the much harder $p_T$ 
spectrum of jet fragmentation.  There is significant enhancement of the baryon/meson ratio in the ridge, quite similar to the
bulk medium (and again in contrast to vacuum jet fragmentation).  Finally the ridge yield appears to scale per trigger particle
as the number of participating nucleons (similar to the bulk medium).  Note that for the 'hard' and 'just right' ridge, the
ridge yield is only of order a one percent modulation of the yield of the bulk medium.  This is not true for the 'soft' ridge
where the yield is a much larger fraction, though the experimental methodology is very different (as we discuss later).

%``Theoretical free for all''  Paul Stankus.  ``Theorists, help us kill your model'',  Brian Cole QM08.

One additional comment regarding the `hard ridge' relates to the question of whether the ridge persists up to the
highest $p_T$ trigger particles.  Results from PHENIX~\cite{phenix_dihadron} show no evidence for the ridge
with trigger particle $p_T$ $>$ 5 GeV/$c$.  However, as noted in the publication, due to the limited pseudo-rapidity acceptance,
if the jet peak is increasing with trigger $p_T$, it may be that the result is not sensitive to the smaller ridge
contribution.  This appears to be the case since STAR preliminary results show a significant ridge yield for 0-10\% central Au-Au
events that appears constant all the way up to $p_{T} ({\rm trigger}) > $ 8 GeV/$c$ with $p_{T} ({\rm associated}) > $ 2 GeV/$c$~\cite{putschke}.

\subsection{Ridge Models}

The various theoretical models can be grouped into two categories.  The first are referred to as
{\bf 'Causation Models'} (i.e. where $A$ (for example a jet parton losing energy) causes $B$ (the ridge)).    The second
are referred to as {\bf 'Auto-Correlation Models'} (i.e. where $C$ causes $A$ and $C$ also causes $B$, such that
$A$ and $B$ are auto-correlated).

Most Causation Models assume that the cause of the 'ridge' is a quenched jet.  Examples include
calculations where the radiation from the jet is broadened
in rapidity due to collective flow~ \cite{armesto_ridge} and 
where the radiation from the jet is broadened by turbulent color fields~\cite{majumder_ridge}.  
However, in both of these cases the ridge has a Gaussian shape 
in pseudo-rapidity with $\sigma_{\Delta \eta} \approx 0.4-0.5$ units.  It is difficult for models 
with quenching parameters that reproduce
the suppression of high $p_T$ hadrons to achieve a broader ridge width.  Note that these models would also provide no
explanation for the 'soft ridge'.  The PHOBOS experiment has shown the 'just right' 
ridge to have a width that extends over four units of pseudo-rapidity~\cite{phobosridge}.  One might
have concerns that this four units in pseudo-rapidity might be quite different in rapidity.  Since PHOBOS includes
associated particles down to $p_T$ = 0.035 GeV/$c$, for $\Delta \eta$ = 4, that would only be $\Delta y \approx $ 1 for protons and
$\Delta y \approx$ 2.5 for pions.  However, the phase space contributions at this lowest $p_T$ are not very large.  In addition,
the STAR preliminary results indicate a width of $\sigma_{\Delta \eta} >$ 1.4~\cite{netrakanti_qm09}, 
and since the $p_T$ associated particles are much higher, pseudo-rapidity and rapidity are much closer.
Thus, these causation models are essentially ruled out as the dominant mechanism for producing the ridge.

\begin{figure}[htb]
\centering
\includegraphics[scale=0.29]{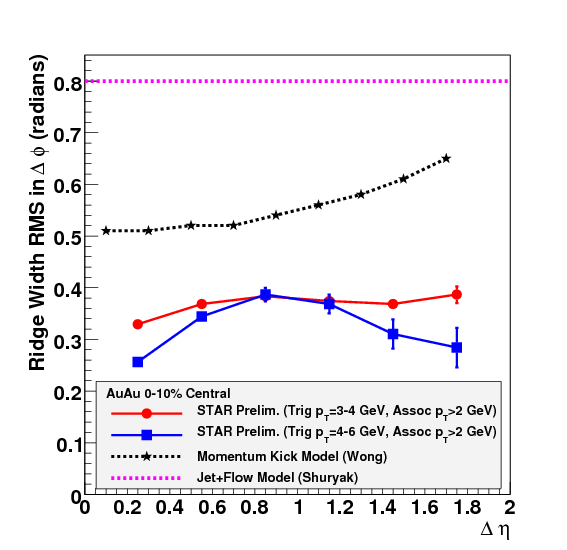}
\includegraphics[scale=0.29]{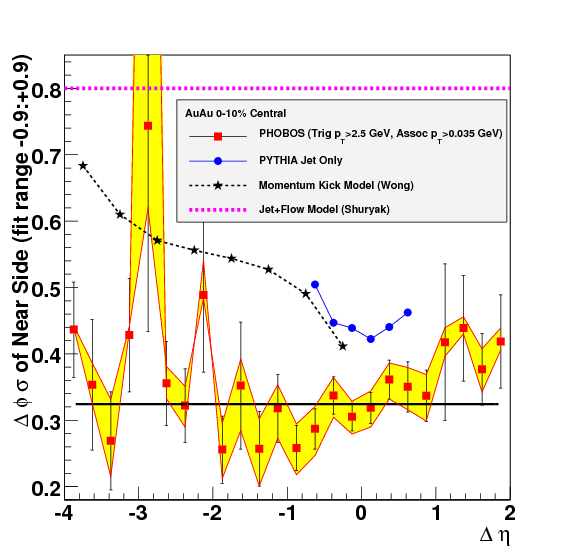}
\caption{(color online) Left (right) panel shows the STAR preliminary (PHOBOS published) data for the ridge width 
in $\Delta \phi$ as a function of $\Delta \eta$.  Shown in comparison are calculations from various
theoretical models described in the text.
}
\label{fig_ridge_theorycomp}
\end{figure}

Another Causation Model is the 'Momentum Kick Model'~\cite{wong1}.  The model includes an effect that is similar to an initial state $k_T$ scatter,
but from the jet parton.  The soft partons are given a momentum kick (given by the parameter $\vec q$), which is tuned
to match the data (where the kick is assumed to be exactly along the direction of the jet parton).  One concern in this
model is that one needs to know the y and $p_T$ distribution of the soft partons, and an arbitrary distribution set could 
essentially match any possible ridge data.  Thus, it is critical to demonstrate that these distributions are not
arbitrary (e.g  by having additional (non-ridge) tests and constraints on them).  With the currently published parameter
set, this model gives a reasonable description of the ridge yield and full extent of $\Delta \eta$ = 4 as shown by the 
PHOBOS data.  One can also compare the width of the ridge in $\Delta \phi$, which has been a somewhat overlooked key
feature of the ridge.  As shown in Figure~\ref{fig_ridge_theorycomp}, this model significantly over-predicts the width of the ridge in
$\Delta \phi$ and has a trend for a wider ridge width at larger rapidity gap.  One can understand this effect by considering the case where
PHOBOS measures a trigger particle at pseudo-rapidity +1.5 and the associated particle
at pseudo-rapidity = -2.5 (to get a gap of 4 units).  Since the momentum kick $\vec{q}$ is always along the trigger parton, some of
the kick is longitudinal (and thus there is less focusing in $\phi$).  
Finally, the Causation Model ~\cite{vlad} postulates that the away-side jet produces a back-splash in the medium.  Thus,
the large range in pseudo-rapidity is caused by the away side jet swing (assuming the initial two partons that
scattered have $x1 >> x2$ or vice versa).  This interesting idea requires a full simulation of the time scales
involved to create the correlation to further test this picture.

\begin{figure}
\centering
\includegraphics[scale=0.29]{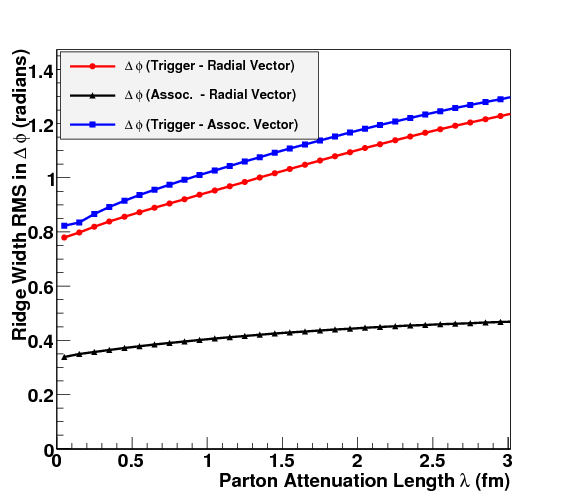}
\includegraphics[scale=0.29]{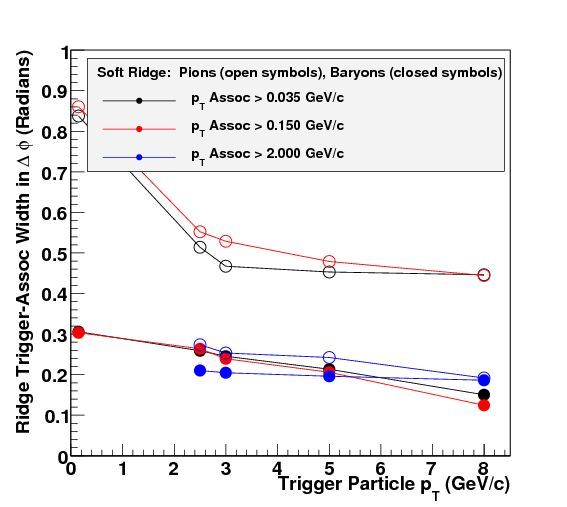}
\caption{(color online) (Left panel) Calculation of the trigger particle - radial vector, associated particle - radial vector, and
trigger and associated particle azimuthal correlation as a function of the jet-quenching attenuation length.
The calculation employs a realistic Glauber geometry for Au-Au central events, and the methodology outlined
in ~\cite{shuryak}.
(Right panel) Calculation results for the ridge width in $\Delta \phi$ as a function of the trigger
particle $p_T$ and for various range of associated particle $p_T$.  The results are shown for pions and protons.
}
\label{fig_ridge_jetinduced}
\end{figure}

%\subsubsection{Auto-Correlation Models}
The second main category of explanations are 'Auto-Correlation Models'.  A common
feature of these models is that there is a local hot spot (in the transverse plane) in the initial collision (i.e. at
the earliest time).  This hot spot expands longitudinally (and thus in rapidity space) and produces both the trigger
and associated particle by different mechanisms.
However, because they originate from a common location
in the transverse plane, they may both be correlated with the radial vector in the overlap region (and thus
auto-correlated into an extended ridge in $\Delta \eta$ and narrow in $\Delta \phi$).  One such calculation is the 'Jet Induced
Ridge' model~\cite{shuryak}.  In this picture there is no specific description of the
longitudinally extended hot spot, and thus there is no quantitative prediction of the width in $\Delta \eta$.  However,
the trigger particle is focused along the transverse radial direction via jet quenching (since the radial vector
represents the shortest path out of the medium).  The associate particle (many units in rapidity away) is focused
via radial flow from the outwardly exploding medium.  We have re-calculated the results from this model and 
find that for any jet quenching attenuation parameter ($\lambda$ which exponentially
suppresses particles traversing a given path length), the narrowest possible ridge in $\Delta \phi$ is approximately
0.8 radians.  As shown in Figure~\ref{fig_ridge_jetinduced}, the trigger particle does not experience 
sufficient focusing to describe the ridge width in $\Delta \phi$.  Also, it is notable that for $\lambda = 1.0$ fm, the single hadron
nuclear modification factor is already $R_{AA} < 0.1$ (significantly lower than published experimental data).  Note
that the ridge is too wide in $\Delta \phi$ both for the STAR and PHOBOS data (as also shown in Figure~\ref{fig_ridge_theorycomp}).

It is striking that the radial flow is much more effective at focusing the associated particle in the above model.
Thus, an alternative model is that both trigger and associated particles are focused by radial flow.  This was
originally proposed in~\cite{voloshin} and is incorporated into the other calculations~\cite{cgc_softridge1, cgc_softridge2}.
However, they also implement a specific description of the hot spot's longitudinal correlation (namely color
glass condensate or 'glasma' flux tubes).  This provides a reasonable description of the 'soft ridge' from the STAR preliminary data.
This model would not have a prediction for the 'hard ridge' case.  At this conference, a hybrid
model was proposed~\cite{moschelli} where focusing by radial flow is only for the 'soft ridge' and at
higher $p_T$ trigger there is an increasing
contribution from 'jet induced ridge' correlations.  This model gives a better description; however, for very high
$p_T$ triggers (e.g. $>$ 8 GeV/$c$) it should predict identical results to the 'Jet Induced ridge' and fail to describe
the experimental data.

We performed a simple calculation, shown in Figure~\ref{fig_ridge_jetinduced}, of the
'ridge width' in $\Delta \phi$ assuming focusing of the trigger and associated particle by radial flow only.  
In the case of radial flow, the higher the $p_T$ of the particles, the larger the implied
radial boost received and thus the more focusing along the radial direction.  This simple calculation
was performed assuming a temperature $T=$140 MeV and a linear boost profile with $\beta_{max} = 0.7$.  The 
exact widths are sensitive to these parameters and the linear boost profile assumption.  However, this type of
model makes a specific prediction, that for moderate $p_T$ trigger and associated particles, if the particles are
protons instead of pions, the focusing is substantially larger.  This prediction needs to be checked experimentally.

A model that takes this soft boosted picture into a full hydrodynamic regime is by Takahashi {\it et al.}~\cite{takahashi}.
In this calculation they have fluctuating event-by-event initial conditions from NEXUS, then followed by
SPHERIO hydrodynamic evolution, and completed with Cooper-Frye hadronization.    In this picture, the longitudinally
extended hot spot (from the underlying NEXUS dynamics) results in a very clear ridge-like structure.  It is notable
that there is an additional jet-like peak, despite the fact that the model has in principle no jets traversing
the hydrodynamic medium.  More precise predictions for these ridge features as a function
of trigger and associated particle $p_T$ are needed as next steps.

%\subsection{Narrow Ridge in $\Delta \phi$ and ZYAM}

\begin{figure}
\centering
\includegraphics[scale=0.25]{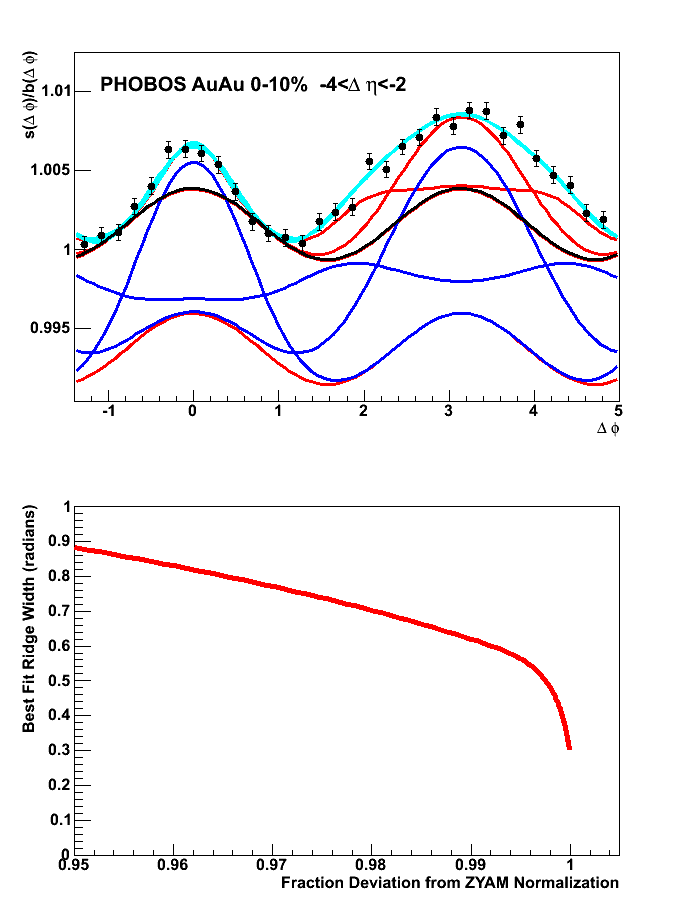}
\includegraphics[scale=0.3]{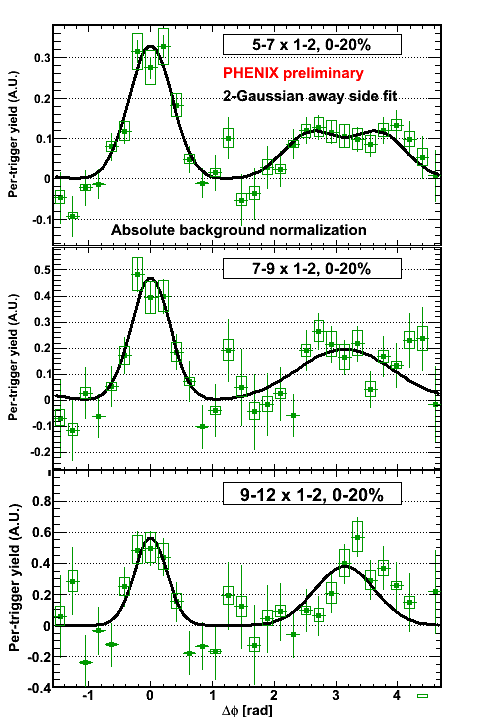}
\caption{(color online) (Left panel) Shown are the PHOBOS published azimuthal correlation function 
(before any ZYAM or other subtraction) 
for the pseudo-rapidity range $\Delta \eta =$ -2.0 : -4.0 (i.e. away from the near-side jet peak).  We have forced
the ridge width to specific values (stepping over a large range) and always find a good fit to the
experimental data given a particular (free parameter) background normalization level (deviating from ZYAM).  
The upper panel shows the experimental data and example fits with a ridge width of $\sigma_{\Delta \phi} \approx$ 
0.3 and 0.6 radians.  The lower panel shows how the normalization level deviates from ZYAM for different forced
ridge widths.  Note that a 1\% deviation in the normalization can increase the ridge $\sigma_{\Delta \phi}$ from 0.3 to 0.6, and a 3\% deviation can increase it to 0.8.  (Right panel) $\pi^{0}$-hadron azimuthal correlations~\cite{adare}.
}
\label{fig_zyam_check}       % Give a unique label
\end{figure}

At the conference, R. Hwa pointed out that the narrow width of the ridge in $\Delta \phi$ could be an artifact
of the ZYAM normalization.  Since one forces a zero point in the correlation function (which turns out to be at
$\Delta \phi \approx 1$), it is difficult to find a width in $\Delta \phi$ greater than 0.3-0.4 units.  
Thus, for the PHOBOS and STAR ridge results, the width values shown
could be under-estimates.  As shown in Figure~\ref{fig_zyam_check}, we have calculated how much the true normalization would have to differ from the
ZYAM normalization in order to accommodate a much wider ridge (utilizing the published PHOBOS raw correlation function data~\cite{phobosridge}).  
We have constrained the ridge width in $\Delta \phi$ to fixed values (stepping over a wide range) and then left as free parameters
the ridge yield, the away side structure (broken into a punch through peak centered at $\pi$ and two shoulder 
peaks centered at $\pi \pm D$), and the $v_2$ modulated background normalization.  We find that for any value of the ridge width
in $\Delta \phi$, one can achieve an equally good description of the experimental data.  As shown in the lower left panel, if there is a 1\% deviation
in the background normalization from ZYAM one can have a width in $\Delta \phi$ of 0.6 units, and for 3\% a width of 0.8 units.  In the
case of the PHOBOS data however, they have also checked that the ZYAM normalization follows an expected scaling
with centrality (which is akin to the absolute normalization technique~\cite{abs_norm}).
Though it deserves additional scrutiny, it appears that even a 1\% normalization difference from ZYAM is unlikely.
The STAR 'hard ridge' results also should be checked with absolute normalization.
Note also that if the normalization is different by 1\%, such that the ridge $\sigma_{\Delta \phi} \approx 0.6$ units (in 
agreement with the 'Momentum Kick Model' for example), the ridge yield would be much larger (then in disagreement
with the 'Momentum Kick Model').  

A point that deserves more attention is that the 'soft ridge' has a width in 
$\sigma_{\Delta \phi} \approx$ 0.9-1.2 units~\cite{daugherity,cgc_softridge1}.  However, this preliminary analysis is not 
done with ZYAM, but fitting all components expected in the correlation function with floating normalizations.  
As the data moves towards publication, a full
quantification of the correlated uncertainties in these parameters is crucial, as they currently have a much 
larger ridge yield and much wider width than the ridge results that utilize ZYAM.
As previously mentioned, the 'soft ridge' yield is very large (a substantial 
fraction of the entire yield), which would also be true if we forced the ridge width to be $>$ 1.0 units for
the PHOBOS correlation.  We note that in the above test we performed with the PHOBOS data, we can get an equally
good fit to the data (i.e. within $\Delta \chi^{2} < 1.0$ from the best fit) for any normalization varying from
ZYAM (even greater than 15\%).  Thus, in that particular test, the fit alone gives essentially no constraint on the
ridge width and yield.

The models with an auto-correlation of soft boosted trigger and associated particles appears
quite compelling.  This physics must be there at some level.  However, the ridge structure appears to persist
even for trigger $p_T$ $>$ 8 GeV/$c$.  This seems to contradict many other pieces of evidence that 8 GeV/$c$ $p_T$ particles
are not predominantly soft thermal particles with a very large radial boost.  
In the hybrid approach~\cite{moschelli}, the ridge should show a significant
broadening at high $p_T$ as one transitions to the 'jet induced' case.  This effect is not seen, though perhaps it
is masked by the ZYAM normalization.  It will be important to do more checks for both the PHOBOS and STAR ridge cases.
Data shown at this conference on three-particle correlations should also be of great interest, though the models
at this point are already challenged by the existing two-particle correlation data and trends.  

\section{Away Side Structure}

Di-hadron azimuthal correlations have also revealed interesting physics on the opposite side of the
trigger particle.  For the case of moderate $p_T$ trigger particles and low to moderate $p_T$ associated
particles, there appears to be a split-peak on the away side.  It has been postulated that this
structure is a mach-cone from a super-sonic parton~\cite{cone,cone2} or gluon Cerenkov radiation or other
medium disturbances.  The STAR experiment has published exciting results on three-particle correlations~\cite{conical}.  It is notable that there are now observations of similar phenomena from CERN-SPS data that challenge
the various theoretical models as a response of the medium (without significant dissipation).
Simplified models can reproduce the main features of the experimental data~\cite{cone2}, but full hydrodynamic calculations of a medium response have not.  Additional attempts within the
context of AdS/CFT are also being made.  

Shown in Figure~\ref{fig_zyam_check} are results that indicate a broad away side correlation for modest 
trigger $p_T$ particles, but no statistically significant indication for higher $p_T$ particles of anything
except a single Gaussian peak.   Most
medium response models should predict a continued response for higher $p_T$ trigger particles, though none
is currently seen.  We note that it is not ruled out that the ``punch-through'' peak (or potentially
tangential emission) may be subsuming the medium response.  This is a key additional experimental
test of the various explanations.

An alternative picture is emerging that proposes that the away-side structure is 
due to auto-correlations of an initial hot-spot with the transverse geometry and flow~\cite{awayauto,takahashi}.
One particular idea relates to strong initial state fluctuations that produce event-by-event 
spatial anisotropies that can translate into surprising momentum anisotropies with various
Fourier coefficients (for example $v_{3}$).  There are also pictures in-between where parton recombination
couples these soft medium particles with those from the initial di-jet partons~\cite{hwa} (relating
both the away-side and ridge structures).
These ideas are getting serious consideration and
need to also be confronted with the full set of experimental observables in a quantitative manner.

\section{Summary}

This exciting area of heavy ion physics has the potential to teach us an enormous amount about
the hot partonic medium created.  To realize this potential, experimentalists will need to continue
to work hard to publish results with fully quantified systematic uncertainties and theorists will
need to continue to learn from cases where data and theory agree and where they disagree.

%% end of main text

\section*{Acknowledgments} % please check/modify, comment out or delete if not needed

We are thankful for many useful and illuminating discussions and people making their data and time
available including Andrew Adare, Sean Gavin, Constantin Loizides, George Moschilli,
Joern Putschke, Lanny Ray, Carlos Salgado, Anne Sickles, Paul Sorensen, Paul Stankus, Jun Takahashi,
Fuqiang Wang, Edward Wenger, C.Y. Wong, Nu Xu, and others.  Also thanks to Glenn Young and
the conference organizers, and thanks to Larry McLerran who offered to be my
'food taster' after my talk.
We acknowledge funding from the Division of Nuclear Physics of the U.S. 
Department of Energy under Grant No. DE-FG02-00ER41152.

 % do not change 

\begin{thebibliography}{00} % do not change 
   
\bibitem{ppg079}A. Adare {\it et al.} (PHENIX), Phys. Rev. C77: 064907 (2008).
\bibitem{ppg080}A. Adare {\it et al.} (PHENIX), Phys. Rev. Lett. 101: 232301 (2008).
\bibitem{star_dihadron}J. Adams {\it et al.} (STAR), Phys. Rev. Lett. 97: 162301 (2006).
\bibitem{zoww} H.Z. Zhang, J.F. Owens, E. Wang, and X.N. Wang, Phys. Rev. Lett. 98, 212301 (2007).
\bibitem{pqm}A. Dainese, C. Loizides, G. Paic, Eur. Phys. J. C38, 461 (2005).
\bibitem{spain_eloss}N. Armesto, M. Cacciari, T. Hirano. J.L. Nagle, C. Salgado, arXiV:0907.0667 [hep-ph].
\bibitem{techqm}$https://wiki.bnl.gov/TECHQM/index.php/Main_Page$
\bibitem{rwei}R. Wei for the PHENIX Collaboration, these proceedings.
\bibitem{putschke}J. Putschke for the STAR Collaboration, Nucl. Phys. A783: 507 (2007), J. Phys. G34: S679 (2007).
\bibitem{daugherity}M. Daugherity for the STAR Collaboration, arXiv:0806.2121.
\bibitem{phobosridge}B. Alver {\it et al.} (PHOBOS), arXiv:0903.2811.
\bibitem{jchenphenix}J. Chen for the PHENIX Collaboration, these proceedings.
\bibitem{phenix_dihadron}A. Adare {\it et al.} (PHENIX), Phys. Rev. C 78: 014901, 2008.
\bibitem{armesto_ridge}N. Armesto, C. Salgado, U. Wiedemann, Phys. Rev. Lett. 93, 232301 (2004).
\bibitem{majumder_ridge}A. Majumder, B. Muller, S. Bass, Phys. Rev. Lett. 99, 042301 (2007).
\bibitem{netrakanti_qm09}P. Netrakanti for the STAR Collaboration, these proceedings.
\bibitem{wong1}C-Y. Wong, Phys. Rev. C 78, 064905 (2008) and references therein.
\bibitem{vlad}V. Pantuev, arXiv:0710.1882.
\bibitem{shuryak}E. Shuryak, arXiv:0706.3531v1.
\bibitem{voloshin}S. Voloshin, nucl-th/0312065v3.
\bibitem{cgc_softridge1}S. Gavin, L. McLerran, G. Moschelli, arXiv:0806.4718v3.
\bibitem{cgc_softridge2}A. Dumitru, Gelis, L. McLerran, R. Venugopalan, arXiv:0804.3858v1.
\bibitem{moschelli}G. Moschelli, these proceedings.
\bibitem{takahashi}J. Takahashi {\it et al.},arXiv:0902.4870v1.
\bibitem{abs_norm}A. Adare, M. McCumber, A. Sickles, manuscript in preparation.
\bibitem{cone}J. Casalderrey-Solana, E.V. Shurayk, D. Teaney, hep-ph/0602183.
\bibitem{cone2}J. Ruppert and T. Renk, arVix:0710.4124.
\bibitem{conical}B.I. Abelev {\it et al.} (STAR), Phys. Rev. Lett. 102: 052302 (2009).
\bibitem{awayauto}P. Sorensen {\it ibib}; C. Pruneau, S. Gavin, S. Voloshin, arXiv:0711.1991; and others.
\bibitem{hwa}R. Hwa, arXiv:0904.2159v1.
\bibitem{adare}A. Adare for the PHENIX Collaboration, Hot Quarks Conference Proceedings.

\end{thebibliography}
\end{document}